**Measuring co-authorship and networking-adjusted scientific impact**

John P.A. Ioannidis[1,2]

[1]Clinical and Molecular Epidemiology Unit, Department of Hygiene and Epidemiology, University of Ioannina School of Medicine and Biomedical Research Institute, Foundation for Research and Technology-Hellas, Ioannina 45110, Greece

[2]Department of Medicine, Tufts University School of Medicine, Boston, MA 02111, USA

Correspondence should be addressed to John P.A. Ioannidis, Professor and Chairman, Department of Hygiene and Epidemiology, University of Ioannina School of Medicine, Ioannina 45110, Greece. E-mail: jioannid@cc.uoi.gr

Word count: 4754; abstract: 247 words; 3 figures; 1 table

**ABSTRACT**

Appraisal of the scientific impact of researchers, teams and institutions with productivity and citation metrics has major repercussions. Funding and promotion of individuals and survival of teams and institutions depend on publications and citations. In this competitive environment, the number of authors per paper is increasing and apparently some co-authors don't satisfy authorship criteria. Listing of individual contributions is still sporadic and also open to manipulation. Metrics are needed to measure the networking intensity for a single scientist or group of scientists accounting for patterns of co-authorship. Here, I define $I_1$ for a single scientist as the number of authors who appear in at least $I_1$ papers of the specific scientist. For a group of scientists or institution, $I_n$ is defined as the number of authors who appear in at least $I_n$ papers that bear the affiliation of the group or institution. $I_1$ depends on the number of papers authored $N_p$. The power exponent $R$ of the relationship between $I_1$ and $N_p$ categorizes scientists as solitary ($R>2.5$), nuclear ($R=2.25-2.5$), networked ($R=2-2.25$), extensively networked ($R=1.75-2$) or collaborators ($R<1.75$). $R$ may be used to adjust for co-authorship networking the citation impact of a scientist. $I_n$ similarly provides a simple measure of the effective networking size to adjust the citation impact of groups or institutions. Empirical data are provided for single scientists and institutions for the proposed metrics. Cautious adoption of adjustments for co-authorship and networking in scientific appraisals may offer incentives for more accountable co-authorship behaviour in published articles.

**INTRODUCTION**

Appraisal of the scientific impact of researchers, teams and institutions using their publication record and citation metrics [1-6] influences career development, funding decisions, and expert and public perceptions about science. The future of single scientists, teams and large institutions increasingly depends on "publish-or-perish" (or "get-cited-or-perish") principles. In this competitive environment, the average number of authors per paper is increasing [7-10]. An increasing portion of papers in influential journals contain very extensive lists of authors. This may reflect in part the welcome advent of more collaborative research efforts. However, probably several co-authors don't satisfy full authorship criteria. Gift (honorary) authorship has been demonstrated repeatedly [11-13]. Ideally, one should know the nature of the individual contributions of each author in each paper and several journals have adopted listing of contributions [14]. Yet, empirical assessments have shown problems also with listing contributions [15,16]. When asked twice about their own contributions, authors have had only modest agreement in their two responses [16].

While it is often difficult to see what an author has truly done in a specific paper, it may be easier and more informative to examine one's overall co-authorship behaviour across one's whole publication record. Current systems of measuring productivity and citation impact for individuals or groups count all papers and all citations the same, regardless of what each author has contributed. Such exercises have acquired strong supporters and have also raised major objections [4-6, 17-20]. For example, the most popular metric currently is the Hirsch $h$ index [1,2], defined as the number of articles (of a scientist, group, or institution) that have received at least $h$ citations each. The original paper describing $h$ [1] is already a most highly-cited article itself with approximately 200 citations received per GoogleScholar. However, neither $h$ nor other similar indices provide information about the co-authorship pattern of a scientist. Two scientists may have the same $h$, but one may have no or only few co-authors in all her papers while the other may be a participant of one or several large collaborations and may have co-authored all her papers with dozens of others.

Similar difficulties appear in appraising teams and institutions [4-6]. Several popular ranking exercises of institutions and universities have received fierce criticism; a key problem is suboptimal accounting for institutional size [4-6]. A larger institution is expected to publish more papers, receive more citations and have higher *h*. Defining the size of a team or institution with administrative data is difficult. Quotas for size would differ enormously depending on whether tenured faculty, tenure-track faculty, associates, post-graduates, pre-graduates, supporting staff, and close collaborators (some not even in the same location) are counted or not.

Here, I propose simple indices that measure the networking intensity, the effective size of a network, for scientists or groups of scientists who co-author papers. These indices are empirically demonstrated with data on authors from the medical sciences and comparatively on authors from physical scientists, where "mega-authorship" papers with hundreds of authors are already commonplace. I also demonstrate the use of these indices for institutions.

**RESULTS**

**Co-authorship networking for single scientists**

For a single scientist, I define $I_1$ as the number of authors who appear in at least $I_1$ papers of that scientist. $I_1$ increases with increasing number of publications $N_p$. With more publications, opportunities arise for having more co-authors and for more papers written in common with each co-author. This relationship can be expressed with a power law $N_p=(I_1)^R$. *R* is calculated as the ratio $log_{10}(N_p)/log_{10}(I_1)$. *R* reflects the co-authorship pattern. With fewer co-authors per paper, for the same $N_p$ the $I_1$ is smaller and *R* larger. For the same $I_1$, *R* increases, when a scientist writes more papers (larger $N_P$) with new or sporadic co-authors who don't contribute to $I_1$; or keeps co-authoring only with a small core of his most common co-authors.

$R$ may be imprecise when the two measures that enter into its calculation, $N_p$ and $I_1$, are small, because then small changes may result in considerable changes in $R$. I recommend to view $R$ very cautiously if $N_p$<30 or $I_1$<4.

$I_1$ values are shown in Figure 1 as a function of the number of papers $N_p$ for highly-cited scientists in Clinical Medicine and Physics according to ISI (ISI highlycited.com. Available at: http://isihighlycited.com Last accessed 2007 December 30). In particular for Physics, the diversity is extreme with $I_1$ values ranging from 6 to 235. The range of $I_1$ for Clinical Medicine highly-cited scientists is 9 to 25. About a third of the examined highly-cited scientists in Physics have $I_1$ above 130. They are all physicists who participate routinely in extremely multi-authored collaborations, mostly in high energy and particle physics. They have written few, if any, papers as first authors, but based on plain citation counts they are among the 250 most influential people in their science. The values of $R$ also range widely from 1.07 to 2.81.

I propose the following classification, based on $R$, to categorize the co-authorship networking of scientists:

1. solitary ($R$>2.5)
2. nuclear ($R$=2.25-2.5)
3. networked ($R$=2-2.25)
4. extensively networked ($R$=1.75-2)
5. collaborator ($R$<1.75)

The proposed cut-offs for $R$ are simply an operational proposal. If $R$ is determined for large numbers of scientists in a specific scientific field, it would also be possible to obtain quintile cut-offs empirically. With increasing numbers of co-authors per paper, typical $R$ values may tend to get lower over time for several scientific fields.

Figure 2 plots the $R$ ratio and the $h$ index for these same scientists. This gives a more complete picture of the performance of a scientist, since it shows not only the citation impact, but also the co-authorship networking.

For authors with common names, it is unlikely that same-name authors would share also co-authors. Therefore, if one inadvertently measures indices for a common name, $N_p$ will increase (accumulation of papers from many authors), but $I_1$ may not increase similarly. This will lead to a spuriously high $R$. For example, for Smith F, we get $N_p$=809, $I_1$=9, $R$=3.05.

One may also wish to adjust the $h$ index for co-authorship networking. I propose an adjustment that would standardize the $h$ index to what its value would have been for a typical "average" networking $R$=2.00 (at the border between networked and extensively networked). To do this, in general one may multiply $h$ by $(R/2)^k$. Unadjusted analyses have $k$=0. With increasing $k$ values, the citation impact of solitary-profile scientists is heightened, while the citation impact for collaborator-profile scientists decreases.

For example, with k=1, $h$ is multiplied by $R/2$ to get $h_{R=2}$. This standardization decreases the $h$ index of the collaborator physicists to a median $h_{R=2}$ of 29 (range 23 to 38) from the original unadjusted median $h$=53 (range 44 to 63). For the other highly-cited physicists in the sample, the median $h_{R=2}$ increases to 71 (range 35 to 128) from the original unadjusted median $h$=61 (range 37 to 116). For the sample of highly-cited scientists in medicine, median $h_{R=2}$ is 86 (range 60 to 167) vs. the original unadjusted median $h$=83 (range 59 to 123). Upward or downward changes for individual scientists are considerable.

$R$ should not be confused with the total number of co-authors of a scientist during his career. Scientists who typically co-author articles with very large established networks of investigators will have low $R$ values because the same co-authors appear again and again in their publications. Conversely, some other scientists may also count cumulatively many co-authors, but these co-authors may be different each time in each paper. The $R$ index will classify such scientists as solitary or nuclear. Paul Erdős, a legend for the number of people he co-authored papers with, is a classic example. In ISI, Erdős has $N_p$=671, $h$=38 and his $I_1$ is only 11, precisely because during his life he kept moving and working with new people each time. For Erdős, one gets $R$=2.71, a most solitary R value. His biographers stress exactly his solitary path where he never really settled to be part of an established network. Per Wikipedia: "He would typically show up at a colleague's doorstep and announce "my brain is open", staying long enough to collaborate on a few papers before moving on a few days later. In many cases, he would ask the current collaborator about whom he (Erdős) should visit next."

Many scientists have relatively stable $R$, i.e. the same networking profile, throughout their career (Figure 3). A solitary scientist may remain solitary throughout his career and a collaborator may retain the collaborator profile over time. There are exceptions to this rule, however, e.g. the physicist Steven Pearton ($N_p$=1300, $h$=61 as of 2007) had $R$=2.50 in 1987 (at the border of nuclear and solitary), and this became $R$=2.06 (networked) in 1997 and $R$=1.94 (extensively networked) in 2007 as he grew larger teams of co-authors in superconductor research.

**Networking for institutions**

For a group of scientists or institution (e.g. a university, hospital, department, team, or research center), I define $I_n$ as the number of authors who appear in at least $I_n$ papers that bear an affiliation of that specific institution. Table 1 shows the $I_n$ values for various institutions for the papers published in a single year (2003) carrying their affiliation. $I_n$ offers a simple measure to approximate the effective networking size of an institution for a given year. $I_n$ increases with increasing number of papers authored with the affiliation of interest.

$I_n$ is susceptible to clustering of extremely multi-authored papers in an institution. The institution-affiliated collaborator authors inflate the top ranks of authors that contribute to $I_n$. Occasionally they may also carry with them some of their collaborators from other universities, further inflating $I_n$. This phenomenon is practically limited to high-energy and particle physics. Excluding such physics papers from calculations considerably reduces $I_n$ for some institutions (Table 1). This corrected value is more representative.

Another artefact can be introduced, if different scientists with the same name in the same institution cluster as the same person and inflate $I_n$, e.g. the Agricultural University of Tokyo spuriously seems to have the same effective networking size ($I_n$=23) as Harvard University. The same problem may arise with any Japanese institution and possibly other national institutions where many names are redundant. Close inspection of the lists of most-prolific scientists, shows this is not an issue for American or European institutions which may also have some Japanese scientists: it is not common to have two same-name prolific scientists in the same foreign institution.

One may similarly define $R$ also for groups and institutions as $R= log_{10}(N_p)/log_{10}(I_n)$, but extra caution is needed. Most institutions take $R$ values around 3. The extremes in Table 1 are 2.60 (National Institute for Human Genome Research [NHGRI]) and 3.24 (Federal University of Rio de Janeiro). These two institutions have the same $I_n$=8, but very different number of published papers in the year 2003 (244 versus 844).

For large institutions, adjustment of citation impact should be performed with $1/I_n$ rather than $R$. Institutions represent a mix of authors with variable types of networking behaviour. Also a single year offers only a snapshot of the career of each author. Overall, $R$ would increase, when there are more authors affiliated with the institution who don't publish enough papers to contribute to $I_n$. The difference in $R$ between NHGRI and Rio de Janeiro is attributed to a much longer tail in Rio de Janeiro of authors publishing <8 papers each but who nevertheless cumulatively publish many papers. This long tail may reflect low productivity of many affiliated authors, or a different mix of scientific fields. For example, NHGRI conducts genetics research (a high-output discipline), while a university includes diverse scientific disciplines, several of which publish few papers per author per year. Disciplines with inherently low productivity don't contribute to $I_n$, i.e. they don't increase the estimated networking size of an institution. This is appropriate, because these disciplines usually also get few citations even for excellent work, since fewer papers are published in their fields. Inactive authors also don't contribute to $I_n$; this is also appropriate, since $I_n$ reflects the *effective* networking size.

The ratio of institutional $h$ over $I_n$, $Q= h/I_n$, may be used as a measure of citation impact adjusting for the effective networking size. For example, the papers published with a University of Texas affiliation in 2003 received in 2003-2007 six times more citations than the papers published with a Tufts University affiliation, but both have $Q$=5.0, suggesting that both institutions produce on average research of similar citation impact. $Q$ values in Table 1 range from 3.1 to 7.1. Lower values are possible: e.g., in the same time period, the National Academy of Sciences of the Ukraine has $Q$=2.3 ($I_n$=9, $h$=21) and University of Panama has $Q$=1.5 ($I_n$=4, $h$=6).

For institutions that focus on common mainstream biomedical and/or physical science disciplines, one may say that $Q$>6 is outstanding, 4.5-6 very good, 3-4.5 good and <3 fair, using a time window of 5-year citation impact (including the year whose published papers are analyzed). However, one should be cautious with simplifications. $Q$ may depend on the mix of disciplines involved in each institution, e.g. Harvard is excellent in Mathematics, but similar analysis on papers carrying Harvard Univ and Dept Math in the same affiliation yields $Q$=3.0.

One may examine also whether smaller teams or sub-institutions with a similar research orientation have similar or dissimilar $Q$ indices. The five major medical centers affiliated with Harvard Medical School (Table 1) have $Q$ indices between 5.7 and 7.1, close to the $Q$ value for Harvard University as a whole (6.7) and Harvard University Medical School ($Q$=7.0). Similarity of $Q$ values occurs despite considerable differences in the size of each medical center ($I_n$ 7 to 19; $I_n$=21 for the medical school).

The appropriate time window for institutional citation impact can be debated and there is no perfect choice [5]. Recent papers may still have not accumulated their complete citation impact, while papers published a long time ago do not reflect the current status of an institution [3]. Increasing the citation window to 10 years (papers published in 1998, citations until end-2007) increases all $Q$ values, but discipline-related differences remain (all Harvard Univ $Q_{10}$=8.4, Harvard Univ SAME Dept Math $Q_{10}$=3.8).

**DISCUSSION**

The proposed indices provide a simple overall impression of co-authoring behaviour. This allows examining whether a scientist has been mostly a participant in large established collaborations vs. a nuclear or solitary investigator. Nuclear or solitary investigators would either work with a core of few colleagues or keep changing collaborators rather than settling within a single large collaborative network. The indices may also convey a sense of how large is the effective networking size of a group or institution. This information may complement and adjust productivity and citation metrics.

Adjustment for co-authorship may correct some of the major limitations of traditional bibliometric indices [17-20]. Not all authorships are created equal, even within the same paper [11]. Ideally, one would like to know explicitly and truthfully the contribution of each author in a scientific paper, but this goal is often not met. Many journals still don't report contributions or report them vaguely. Authorship position (first, senior, middle position) may offer hints on contributions, but this varies across scientific micro- and macro-environments with divergent authorship cultures [21-24]. Moreover, even if one gives extra bonus to first authorships, this does not solve the challenge of sorting contributions of authors in other positions.

Alternative quantitative approaches for adjusting for co-authorship may also be considered [20]. One may adjust citations for single papers, e.g. dividing the number of citations in each paper by the number of authors and using adjusted citation counts to generate total adjusted-citation counts or respectively normalized $h$ indices, as performed automatically by the Publish or Perish software ($h_{I,norm}$ index) (www.harzing.com). However, sometimes this may be a very stringent adjustment. For example, the 2001 *Nature* paper on the initial sequencing and analysis of the human genome received 5968 citations by the end of 2007, but included 244 authors, thus each author of this truly landmark paper would get credit for only 5968/244≈24 citations. Similarly, mega-authored physics papers would typically give credit for <1 citation to each author; collaborator-profile most-cited physicists may then be re-classified as being among the *least*-cited physicists. Another option is to adjust citations in a paper differently depending on the position of each author, e.g. the first author may get credit for the full number of citations, while the second and the last author may get credit for half, the third may get credit for a third of the citations and so forth. However, it is difficult to reach consensus on what adjustment would be appropriate across different papers and disciplines. Many multi-authored papers simply list authors alphabetically without any connotation of relative contribution in the presented order. Moreover, such complex adjustments would be computationally cumbersome. Conversely, an advantage for the indices that I propose is their easy computation. ISI Web of Science allows the routine automated listing of all authors in a set of papers according to diminishing number of papers to which they have participated. The set of papers can be defined based on the name of the author or institution of interest. Many authors or institutions may be appraised rapidly.

With mounting pressure to publish-or-perish, more authors are squeezed in the same manuscript [7-10,25,26], while the number of papers over the last 35 years has simply grown at the same pace as the number of scientists working in each field [27]. Therefore, the publication and citation record of each researcher becomes inflated primarily by the inclusion of more co-authors per paper, not by taking the lead in more original work. When papers count the same in bibliometric indices, no matter if single-authored, first-authored, or co-authored with hundreds of others, investigators may be willing to be more lenient in including more co-authors. This may even pay off by reciprocal inclusion in each other's papers. Mutually enhancing collaboration may then regress into paper trading. Moreover, the continuous funding and survival of collaborations often depends on the CVs of the leaders; extensive gift authorship is suspected for some influential chairmen [28]. The danger is major, if unscrupulous teams that practice mutual gift authorship extensively [29] make their unscrupulous members more competitive against scientists with more demanding authorship standards.

Increasing number of authors over time does not reflect only increased work that needs to be done per paper. An evaluation focusing on studies with similar design has witnessed a significant increase over time in the number of authors required to run a similar study [7]. The proposed indices may be helpful in addressing this trend of inflated co-authorship, by providing information about the networking pattern of each scientist. While in medicine coalitions of authors are not yet as large as those observed in high-energy physics [27,30], an increasing number of collaborative articles in prestigious journals have many dozens or even several hundreds of authors. Systematic gift authorship or other unsound practices for inflating CVs (e.g. salami publications [31]) would show themselves as more extensive networking in the indices that I propose. At the institutional level, excessive publications and mounting numbers of authors per paper similarly make an institution look larger in effective networking size and will decrease its adjusted scientific impact ($Q$). Adoption of metrics that measure and adjust for co-authorship may offer a disincentive against poor authorship practices. Of course, a "collaborator" profile should not be taken to mean gift authorship. However, authors may be more accountable about who will co-author a manuscript, if they know that inclusion of more authors will decrease their own estimated scientific performance.

Conversely to gift authorship, the opposite trend, ghost authorship, typically occurs when corporate authors write manuscripts for academics and the real authors don't appear on the manuscript [32]. The proposed indices would not be able to pick the presence of ghost authors in specific manuscripts. However, typically these manuscripts also have gift authors. Therefore, the same inflation of networking indices may be observed for these scientists.

The proposed metrics should not discourage true collaboration that serves the needs of science rather than gift authorship. As discussed above, the proposed metrics may downgrade the citation impact of collaborator-profile scientists only modestly, while other methods such as adjustment of citations per number of authors almost totally eliminate their citation impact. Besides established networks producing routinely high numbers of papers, new cross-field collaborations at the margin of disciplines and in newly developing fields are particularly helpful. Such collaborations increase mostly the number of new co-authors in one's CV. This does not affect $I_1$, therefore $R$ decreases and the impact of one's work becomes even more prominent when adjusted by $R$. Therefore, the proposed approach would give more credit to scientists who can create and be involved in new cross-field collaborations.

The networking profile of a scientist may also be examined as it evolves over time. Large-scale evaluations may examine whether indeed most scientists have the same profile throughout their career. A few scientists may also exhibit a mixed networking behavior, e.g. they may have their own solitary methodological work, but they may also be involved as participants in established large collaborations. If their involvement in large collaborations exceeds a certain level, they may get the label of "collaborator", even though they may have also a strong track record in solitary and nuclear work. Such scientists would also have a high $h_{I,norm}$ in contrast to the typical "collaborator" whose impact is almost annihilated in the $h_{I,norm}$ metric.

Some additional limitations should be discussed. The number of indexed papers and citations depend on the database used [33] (e.g. ISI, Scopus, or GoogleScholar). For some fields, specific databases have deficient coverage, e.g. ISI has imperfect coverage in Economics, Computing or Engineering. Citation counts also are affected by entry errors [34], but the impact is small on indices such as $h$, $I_1$ and $I_n$ that are roughly proportional on logarithm of counts [1]. Entries for scientists with the same exact name should be resolved. Also, the proposed indices can be tenuous when $N_p$ is small. Finally, comparisons of scientists and institutions can be misleading, when they work on fields with different citation densities [35].

Automated indices should not replace critical scientific thinking and careful multifaceted evaluation of excellence. However, even with the best intentions, peer assessment may be subjective and occasionally even clearly partial. Moreover, the proposed metrics should not offer an alibi towards lack of transparency about author contributions. We should encourage the complete, transparent, and just communication of contributions for each scientist participating in any project that results in a written manuscript. Nevertheless, automated measures of performance have already made a sweeping presence across scientific fields and they are probably here to stay, so we need to find ways to improve them. Absent a perfect, transparent world on who has done what, at a minimum the proposed co-authorship and networking indices can offer some clues about how each author is networking in publishing scientific work.

**MATERIALS AND METHODS**

**Definitions**

For a single scientist, $I_1$ is defined as the number of authors who appear in at least $I_1$ papers of that scientist. For a group of scientists or institution (e.g. a university, hospital, department, or research center), $I_n$ is defined as the number of authors who appear in at least $I_n$ papers that bear an affiliation of that specific institution.

Calculation of $I_1$ requires ranking co-authors in order of decreasing number of papers that they have co-authored with the specific scientist. Calculation of $I_n$ requires ranking authors in order of decreasing number of papers they have authored where the affiliation of the specific institution is involved. The decreasing-count ranking is conceptually similar to the process used to calculate the Hirsch *h* index for citations [1,2].

**Database of highly cited scientists**

ISI Web of Knowledge includes a list of most-cited scientists in each scientific field based on the number of ISI citations they have received in the period 1981-1999. Approximately 250-300 scientists are included per scientific field. For this analysis, I ordered the scientists in the fields of Clinical Medicine and Physics alphabetically and selected every tenth name for further analysis. Evaluation of publication records and citations was performed in ISI Web of Science for the entire career of each scientist until December 2007. Filters were used to exclude from all analyses meeting abstracts, corrections, and art items since they would increase the count of papers and co-authors without increasing citations perceptibly. Subject category filters were used, when needed, for disambiguation of same-name scientists. Four Japanese scientists and 4 non-Japanese scientists with very common names were not analyzed as it might not be possible to disambiguate with sufficient accuracy which papers were theirs and which belonged to synonymous scientists working in the same scientific field.

Analyses of $I_1$ and *R* indices for the earlier phases of each scientist's career censored the publications of each analyzed scientist at the end of 1987 and 1997, respectively. If a previously solitary scientist starts publishing many papers with many co-authors and the same co-authors are involved in many papers, *R* may gradually decrease. *R* may increase if the opposite scenario occurs (networked scientist becoming solitary). However, if a scientist has already been a common collaborator in many extremely multi-authored papers, *R* will not increase a lot, if he switches to publishing in solitary mode: $I_l$ is already very high and the career-wide $I_l$ can never decrease. This solitary switch would be better captured if the specific period, rather than the whole career is considered.

Previous adjustments of scientific citation impact for co-authorship have considered adjusting the *h* index by the number of authors in the *h* top-cited papers [20]. However, these top-cited papers are not necessarily a large enough or representative sample of the researcher's corpus and the number of authors can be highly susceptible to a few extreme values. The same susceptibility occurs for the total or average number of authors when all articles published by a scientist are considered. For example, an author who writes mostly papers with 2-3 co-authors may have a grossly inflated total or average, if he writes 2 papers with 200 co-authors in each. Distributions of numbers of authors are often far from Gaussian. The median number of authors also does not capture the spread of the distribution. Similar to the *h* index, $I_l$ and $I_n$ has the advantage of being robust to the influence of sporadic papers with extreme counts. Moreover, there is no automated rapid approach currently to record and analyze the number of authors in a set of papers. For large collections of articles the time and effort would be prohibitive.

**Database of institutions**

The analyzed institutions have been selected for evaluation based on the numbers of citations that papers with each institution address have received in the last decade according to Thompson Web of Knowledge Essential Science Indicators (ESI) module (available with subscription from Thomson Scientific Web of Knowledge). Data were collected for the 6 most cited institutions, for the most-cited Canadian, Japanese, and European universities (Univ Toronto [rank 16], Univ Tokyo [rank 13], Univ Cambridge [rank 18], respectively), as well as for systematically sampled institutions that have the ranks 51, 101, 151, 201, 251, 301, 351,401, 451, 501, 551, 601, 651, and 701 for number of citations in the same database. Also data were collected for the 5 major academic medical centers affiliated with Harvard Medical School (those with largest number of published papers in 2003 among Harvard-affiliated hospitals) and for the Department of Mathematics (Dept Math Harvard Univ) and Harvard Medical School. For the main analysis, published articles in the year 2003 were considered and their citation impact traced for the 5-year window start 2003-end 2007.

It is expected that some scientist names and institutional affiliations (in the range of 10% based on detailed analysis of samples of ISI records) may have been miscoded in the ISI databases, and this would slightly underestimate $h$, $N_p$, $I_1$, and $I_n$.

For institutional impact, Kinney has recently proposed [36] that the ratio of the institutional $h$ index by $(N_p)^{0.4}$ characterizes the scientific work quality of an institution. The adjustment by $1/I_n$ that I propose follows the same principle, but there are also some differences. When scientific disciplines have inherently low productivity per author and receive few citations, adjustment by $1/I_n$ will not penalize institutions that foster such disciplines, while the cumulative number of papers may increase considerably. Conversely, adjustment by $1/I_n$ does not penalize an institution as much as $1/(N_p)^{0.4}$ when there are many low-productivity scientists in the institution publishing few papers each and the cumulative number of their papers is substantial.

**Disclosures:** none

**Funding:** none

**FIGURE LEGENDS**

**Figure 1:** $I_1$ values as a function of the number of papers $N_p$ for selected highly-cited scientists in Clinical Medicine and Physics. Both axes are in log-10 scale.

**Figure 2:** Citation $h$ index as a function of $R$ for the scientists of Figure 1.

**Figure 3:** Evolution of $R$ over time as function of the number of papers $N_p$ for selected scientists with different networking profiles.

**Table 1: Networking size and citation impact for various institutions**

| Institution | ESI ranking | $I_n$ ($I_n$ with physics) | $h$ ($h$ with physics) | $R$ | $Q$ |
|---|---|---|---|---|---|
| Harvard Univ | 1 | 23 (26) | 155 (157) | 2.92 | 6.7 |
| Univ Texas | 2 | 24 (29) | 121 (122) | 2.90 | 5.0 |
| Max Planck | 3 | 19 (24) | 103 (108) | 2.93 | 5.4 |
| Johns Hopkins Univ | 4 | 18 (18) | 111 (111) | 2.94 | 6.2 |
| Stanford Univ | 5 | 16 (26) | 108 (112) | 3.05 | 6.8 |
| Univ Washington | 6 | 16 (17) | 115 (115) | 3.14 | 7.2 |
| Univ Tokyo | 13 | 36 (47)* | 88 (92) | * | * |
| Univ Toronto | 16 | 18 (19) | 92 (92) | 3.00 | 5.1 |
| Univ Cambridge | 18 | 16 (26) | 88 (90) | 3.08 | 5.5 |
| Univ British Columbia | 51 | 16 (25) | 70 (71) | 2.92 | 4.4 |
| Tufts Univ | 101 | 13 (13) | 65 (65) | 2.87 | 5.0 |
| Mt Sinai Sch Med | 151 | 13 (13) | 62 (62) | 2.81 | 4.8 |
| Beth Israel Deaconess Med Ctr | 201 | 12 (12) | 80 (80) | 2.87 | 6.7 |
| Med Coll Wisconsin | 251 | 12 (12) | 52 (52) | 2.74 | 4.3 |
| Univ Grenoble 1 | 301 | 8 (14) | 33 (33) | 3.10 | 4.1 |
| NHGRI | 351 | 8 (8) | 46 (46) | 2.60 | 5.8 |
| Charles Univ | 401 | 9 (14) | 32 (34) | 3.22 | 3.6 |
| Univ Fed Rio de Janeiro | 451 | 8 (14) | 25 (26) | 3.24 | 3.1 |
| New York Med Coll | 501 | 9 (9) | 36 (36) | 2.73 | 4.0 |
| Oklahoma State Univ | 551 | 8 (8) | 33 (33) | 3.16 | 4.1 |
| Montana State Univ | 601 | 7 (7) | 33 (33) | 3.16 | 4.7 |
| Tokyo Univ Agr & Technol | 651 | 23 (23)* | 29 (29) | * | * |
| Princess Margaret Hosp | 701 | 8 (8) | 42 (42) | 2.91 | 5.3 |
| Brigham & Womens Hosp | 46 | 19 (19) | 109 (109) | 2.64 | 5.7 |
| Massachusetts Gen Hosp | 39 | 16 (16) | 105 (105) | 2.84 | 6.6 |
| Childrens Hosp (SAME Harvard Univ) | Not listed | 7 (7) | 51 (51) | 2.99 | 7.3 |
| Dana Farber Canc Inst | 165 | 13 (13) | 87 (87) | 2.65 | 6.7 |

NHGRI: National Human Genome Research Institute. Data on $I_n$, $h$, $R$, and $Q$ are based on papers published in 2003 and their citation impact in the 5-year window 2003-2007. Extremely multi-authored physics papers are excluded using subject category filters. Essential Science Indicators (ESI) ranking is automatically generated by Essential Science Indicators module of ISI Web of Knowledge based on citations to papers published in 1997-2007. * considered unreliable due to common Japanese names (artifact)

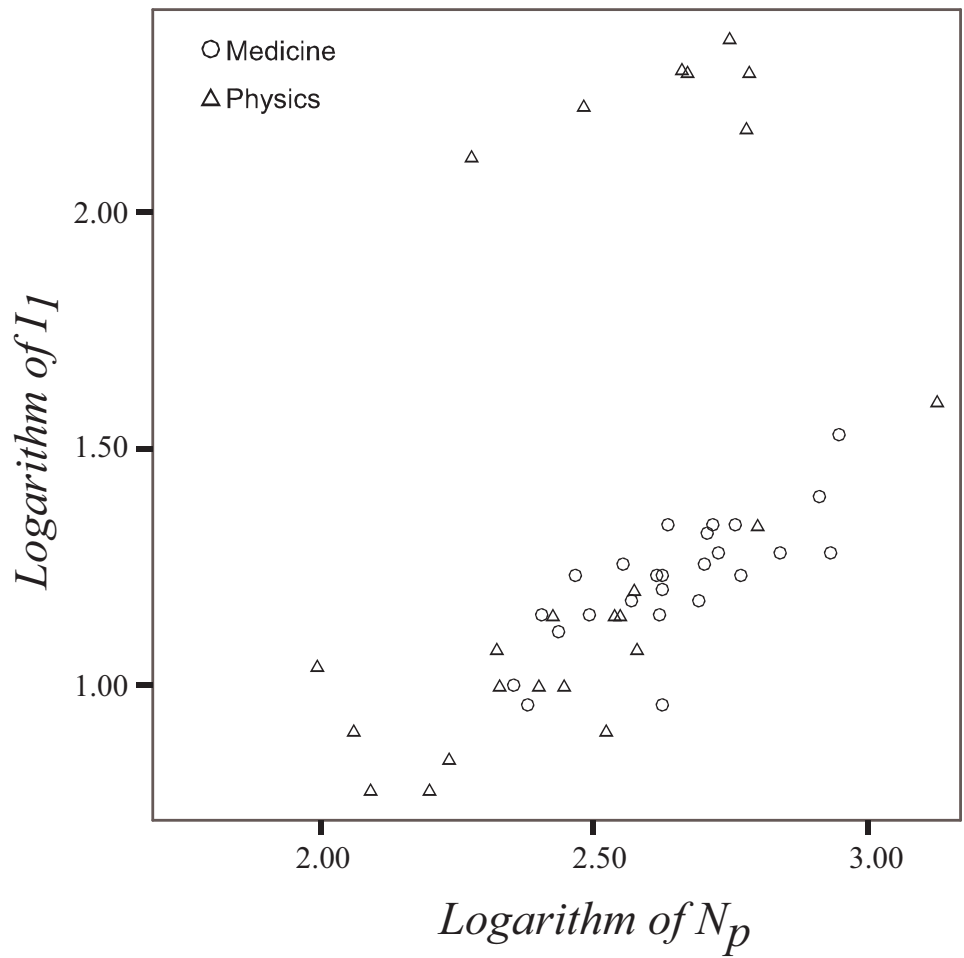

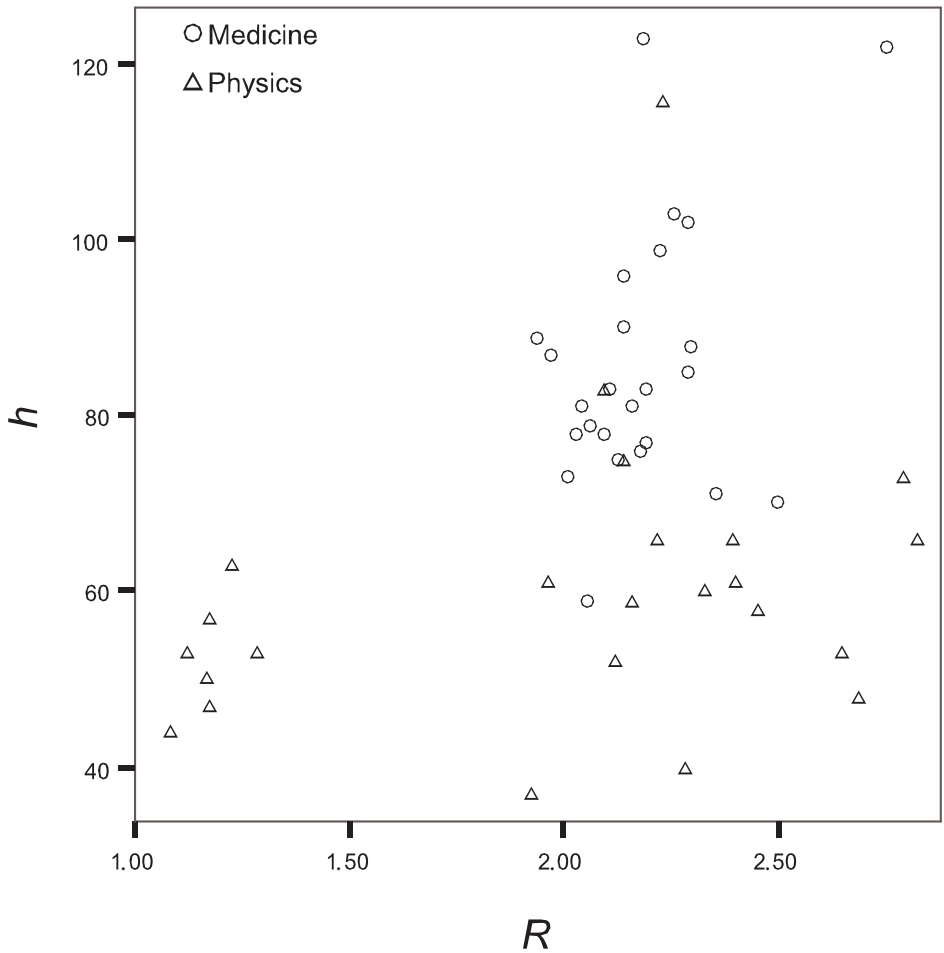

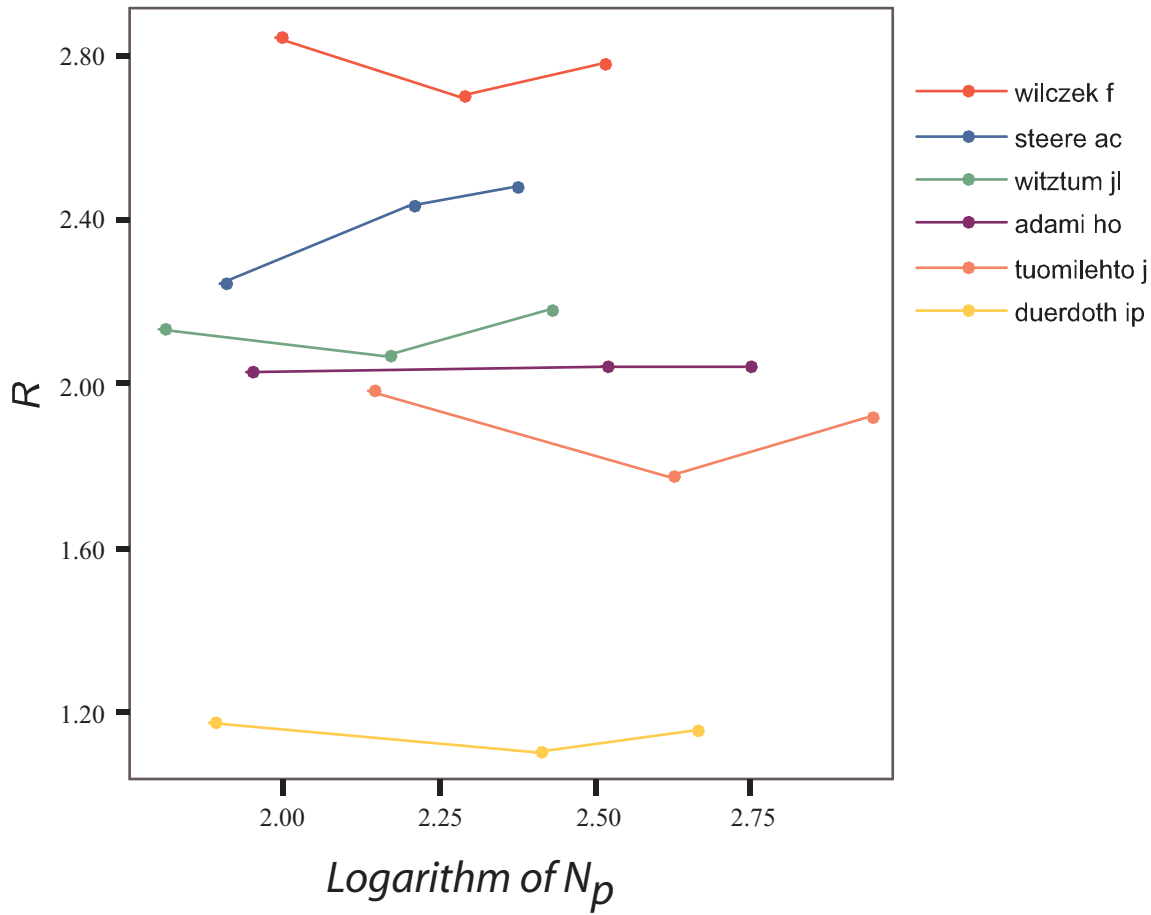